\documentclass[twocolumn]{aastex62}

\begin{document}

\title{Fragmentation in a Primordial Accretion Flow}

\author{Wei-Ting Liao}
\affil{Department of Astronomy, University of Illinois, 1002 West Green Street, Urbana, IL 61801, USA}
\email{wliao10@illinois.edu}

\author{Matthew Turk}
\affil{School of Information Sciences, University of Illinois, 501 East Daniel Street, Champaign, IL 61820, USA}
\affil{Department of Astronomy, University of Illinois, 1002 West Green Street, Urbana, IL 61801, USA}

\begin{abstract}
Under rapid cooling from molecular hydrogen, the accretion disks around Population III (PopIII) stars are believed to fragment, resulting in multiple accreting cores.
In this paper, we build a theoretical framework for calculating the optical depth of H$_2$ ro-vibrational line cooling based on the vertical structure in these accretion disks.
Applying this physically motivated prescription for the optical depth, we find that cooling in the inner disk with $r \lesssim 10 {\rm\ AU}$ is attenuated significantly as a result of high surface density; $PdV$ heating becomes more efficient than cooling, which prevents fragmentation in the inner disk. Despite this, cooling becomes dynamically important in the outer disk, favoring fragmentation. We argue that most of the resultant fragments are initially at the outer disk, and that
any surviving fragment has to migrate slower than the disk-scale photo-evaporation process. Since type I migration is fast, migration slows down as a result of gap-opening in the disk structure. 
Two possible processes for gap-opening are studied: (1) through a massive, densely-cored ($\rho \gtrsim 10^{-8} {\rm\ g\ cm^{-3}}$) clump able to radiate away the excess gravitational potential energy, and (2) through a fast-growing central star, with $\dot{M} \gtrsim 2 \times 10^{-3} \, M_\odot {\rm\ yr^{-1}}$, whose gravity dominates the star-disk system and favors gap opening.
\end{abstract}


\section{Introduction}
Population III (PopIII) stars serve as engines of cosmic evolution; metals released during the end of their life initiate cosmic evolution and transition to the star formation processes observed in stars and galaxies today. 
In a $\Lambda$CDM cosmology, PopIII stars have been believed to primarily consist of massive, low-multiplicity objects with $M_* \gtrsim 100 M_\odot$  and with high accretion rates $\dot{M} \sim 10^{-3} M_\odot {\rm\ yr^{-1}}$ \citep{Abel+02, Bromm+02, Tan+04, McKee+08}.
However, more recent simulations continue to follow the subsequent evolution after the formation of the central protostar.
Subsequent studies have shown that the accretion flow is prone to fragmentation \citep{Turk+09, Stacy+10, Clark+11, Greif+11a}, 
and photo-evaporation ultimately turns off the accretion \citep{McKee+08, Hosokawa+11}. 
However, the mass of the central star and the length scale and frequency at which fragmentation occurs vary across different simulations (see, e.g., \cite{Greif+11a}, \cite{Hirano+14}, \cite{Stacy+16}, \cite{Susa+19}). 

The length-scales and time-scales of fragmentation are key components in determining the mass of PopIII stars, as well as establishing links between numerical models and observations, including of low-metallicity stars in the galaxy today. 
Fragmentation assembles the initial mass function (IMF) for PopIII stars, and the high-order interaction between fragments scatters off few low-mass clumps \citep{Stacy+16}. 
If local metal-poor stars carry the the imprint of primordial metal enrichment process, their metal-abundance patterns should reflect that expected from the IMF of simulated PopIII stars (eg. \cite{Aoki+14}; \cite{Ishigaki+18}).
Furthermore, if inward migration happens frequently, the resulting episodic accretion would delay the photo-evaporation process and allow a $M_* \gtrsim 250 M_\odot$ to form \citep{Hosokawa+16}. 
High star formation efficiency releases significant amount of Ly$\alpha$ photon that would lead to an early 21cm absorption trough, such as that observed at $z \sim 17$ in the EDGES results \citep{Bowman+18, Schauer+19}. 
A high-mass IMF also allows the possibility of a permeated radio background in the early Universe and provides an explanation for the unexpected 21cm absorption depth \citep{Feng+18, Ewall-Wice+18}. 

In this work, we study the process of fragmentation and the survival conditions for those fragments during the accretion phase onto metal-free primordial disks. 
We first review the cooling processes in the primordial gas and derive an approximate cooling function for H$_2$ ro-vibrational transitions in an accretion disk geometry in S\ref{S: cooling}. 
With this adopted optical depth in disk geometry, the resulting conditions for fragmentation are then calculated in S\ref{S: fragmentation}, with an estimation on the subsequent migration rate in S\ref{S: migration}. We study the clump survival condition in S\ref{S: clump}, and conclude in S\ref{S: conclusion}.

\section{Cooling} 
\label{S: cooling}
\subsection{Cooling through H$_2$}

Radiative cooling, particularly in low-metallicity gas, is the second most influential process (after gravity) in the hydrodynamics governing the early phases of star formation.
In a minihalo with $M \lesssim 10^6 M_\odot$, the virial temperature is lower than $10^4 {\rm\ K}$, just below the lowest energy level of atomic hydrogen;  as a result,  molecular hydrogen is the most efficient coolant through ro-vibrational line emission, and thus the most relevant radiative process in star formation process in the early Universe \citep{Barkana+01}. 

During the initial cloud collapsing phase, H$_2$ is collisionally excited, and de-excited through radiative emission. Collisional de-excitation then takes over at $n_{cr} = 10^4 {\rm\ cm^{-3}}$, and the cooling rate becomes linearly proportional to the molecular hydrogen number density until $n \gtrsim 10^{10} {\rm\ cm^{-3}}$, which is roughly the density at which H$_2$ line cooling turns optically thick. 
At even higher density, collision-induced emission (CIE) becomes feasible, as interactions between H$_2$ with another H$_2$ or other species (e.g. H and He) are possible through a dipole transition. And, CIE cooling becomes more efficient than H$_2$ line cooling at $n \sim 10^{15} {\rm\ cm^{-3}}$. 

Following the evolution of primordial protostellar systems requires modeling of both H$_2$ line cooing and CIE cooling.
Although CIE most likely stays optically thin\footnote{At $T \approx 2000 {\rm\ K}$, the Rosseland mean opacity is around $10^{-5}{\rm\ cm^2\ g^{-1}}$ for $\rho = 10^{-8} {\rm\ g\ cm^{-3}}$ \citep{Mayer+05}. However, Rosseland mean opacity increases fast with temperature. If in any case the gas goes above $6000 {\rm\ K}$, e.g. a hot atmosphere, we should expect to have an optically thick condition.}, H$_2$ line cooling becomes optically thick at $n \sim 10^{10} {\rm\ cm^{-3}}$. 
Modeling optically thick H$_2$ line cooling requires modeling radiative transfer in lines, which is known to be both computationally expensive and highly memory intensive \citep{Greif+14}; thus, several substitute methods have been developed.
Under the assumption of spherical symmetry, \cite{Ripamonti+04} calculate the optical depth for H$_2$ line cooling using an isothermal Bonner-Ebert sphere, and provide a fitting function. 
However, soon after the central protostar forms, angular momentum conservation leads the surrounding gas to settle into a flattened disk geometry, at which point the assumption of spherical symmetry is no longer valid.
At this point, a common alternative is to apply a Sobolev approximation \citep{Yoshida+06, Hartwig+15}.

Doppler shifted photons are more likely to ``escape" within damping wings that are at wavelengths surrounding the line center; utilizing this, \cite{Sobolev} approximates the Doppler effect using the velocity gradient as a proxy for the Doppler shifts.
Earlier attempt applying Sobolev approximation in the primordial stellar systems found that the resultant central star would be massive ($M_* \sim 100 M_\odot$) and that the proto-stellar system is resistant to fragmentation \citep{Yoshida+06}. Later simulations were -- critically -- able to evolve the system longer and better resolve the turbulence in the inflow, both of which are critical for quantifying the long-term impact of the Sobolev approximation. 
Fragmentation in simulations of these systems is common, if not ubiquitous (e.g. \cite{Stacy+10}, \cite{Clark+11}, \cite{Greif+11a}, \cite{Greif+12}, to list a few). 
In \citep{Greif+14}, the authors establish through radiative line transfer hydrodynamic calculations that the Sobolev approximation could over estimate the photon escape fraction by an order of magnitude when turbulence is fully-resolved \citep{Greif+14}, and thus the utility of the Sobolev approximation as an approximation in these cases becomes less clear.
In this work, we examine theoretical approximations for the optical depth of H$_2$ line-cooling in disk geometries, and attempt to quantify its resulting role in fragmentation of metal-free accretion disks.

\subsection{Optical Depth of H$_2$ Line Cooling}
Here, we estimate photon escape fraction, $f_{\rm esc}$, for a primordial accretion disk based on the vertical structure of the disk.
For a vertically isentropic accretion disk, the density and temperature profiles are given by
\begin{eqnarray}
T(\eta) &=& T_0  \left( 1 - \eta^2 \right)  \, , \\
\rho(\eta) &=& \rho_0  \left( 1 - \eta ^2 \right)^{\frac{1}{\gamma - 1}} \, .
\end{eqnarray}
$T_0$ and $\rho_0$ are temperature and density at the disk's midplane, $\eta \equiv h/h_0$ is a new disk coordinate in the vertical direction, $h \equiv z / H$, $h_0 = \sqrt{5}$ for a disk with $\gamma = 1.4$, $H \equiv c_{s, 0} / \Omega$ is the disk scale height, $c_{s, 0}$ is the sound speed at midplane, and $\Omega = \sqrt{GM_*/r^3}$.
We note that these are self-similar profiles, where density/temperature field from different radius can be collapsed to the same curve by applying a transformation equivalent to stretching the disk scale height. 
In this new disk coordinate $\eta$, the midplane corresponds to $\eta = 0$ and $\eta \rightarrow 1$ as the coordinate approaches the disk's surface.

In this new coordinate system, we seek to derive a functional form of the optical depth as it varies with $\eta$.
We set the disk surface to be where $T(\eta_s) = 0.1 \, T_0$, which leads to $\eta_s = 0.95$. 
Since density above the disk's surface will provide negligible contribution to the column density, we define the optical depth $\tau(\eta_s) \approx 0$. 
The optical depth can then be written as
\begin{eqnarray}
\label{eq: eta_formula}
\tau ( \eta  ) 
&=& H h_0  \int_{\eta '= \eta_s}^{\eta '=\eta} \widehat{\alpha} (T) \, n_{H_2} (\eta ') \, d \eta '  \\ \nonumber
&=& \left[  \frac{0.76 h_0 \widehat{\alpha}_0 }{m_{\rm H_2}} f_{\rm H_2} \rho_0 H   \right]
\int_{\eta '= \eta_s}^{\eta '=\eta}  \frac{\widehat{\alpha} (T)}{\widehat{\alpha}_0} ( 1 - {\eta ' } ^{5/2} )^{1/(\gamma - 1)} d \eta ' \\ \nonumber
& \equiv & \left[ 76 \,
f_{\rm H_2} 
\left( \frac{\rho_0}{10^{-10} {\rm\ g\ cm^{-3}}} \right) 
\left( \frac{r}{10 {\rm\ AU}} \right)  \left( \frac{H}{r} \right) 
\right] \, \widehat{\tau} \left( T_0, \eta \right)
\, ,
\end{eqnarray}
where $n_{\rm H_2}(\eta) \approx 0.76 \, \rho(\eta) \, f_{\rm H_2} / m_{\rm H_2}$, $f_{\rm H_2}$ is the mass fraction of molecular hydrogen, $\widehat{\alpha}$ is the frequency averaged absorption cross section and is defined as
\begin{eqnarray}
\label{eq: alpha_formula}
\widehat{\alpha} \equiv \frac{ \int \widehat{\alpha}_{\nu} I_{\nu} \, d \nu }{ \int I_{\nu} \, d \nu} \, ,
\end{eqnarray}
and $\widehat{\alpha}_0$ is a normalization factor and is chosen to be $10^{-26} {\rm cm^2}$ throughout this work. 
In the above equation, $I_{\nu}$  is the ro-vibrational emission from molecular hydrogen.

In order to calculate $\widehat{\alpha}$, we need the absorption coefficient for each ro-vibrational transition, $\widehat{\alpha}_{\nu}$. 
The absorption coefficient from an upper energy state ($u$) to a lower energy state ($l$) is 
\begin{eqnarray}
\widehat{\alpha}_{\nu} (T) = \frac{ E_{ul}}{4 \pi} \frac{n_l}{n_{\rm H_2}} B_{lu} 
\left[ 1 - \exp \left( \frac{ E_{ul}}{k_B T} \right)  \right] \phi(\nu) \, ,
\end{eqnarray}  
where $E_{ul}$ is the energy difference between two energy states, $B_{ul}$ is the Einstein absorption coefficient, $n_l$ is the molecular number density at the lower energy state, and $\phi(\nu)$ is the line profile (as a function of frequency). 
According to previous simulation results, turbulence is the primordial gas is generally subthermal \citep{Abel+02}. The line profile $\phi(\nu)$ is thus commonly approximated as
\begin{eqnarray}
\phi(\nu) = \frac{1}{ \Delta \nu_D \sqrt{\pi} } 
\exp \left[ - \frac{ \left( \nu - \nu_0 \right)^2 }{ { \Delta \nu_D } ^2 }   \right] \, ,
\end{eqnarray}
which accounts for the thermal broadening and $\Delta \nu_D = \left( \nu_0 / c \right) \sqrt{k_B T / m_H}$, with $\nu_0$ being the frequency at the line center. 

\begin{figure}
\epsscale{1.2}
\plotone{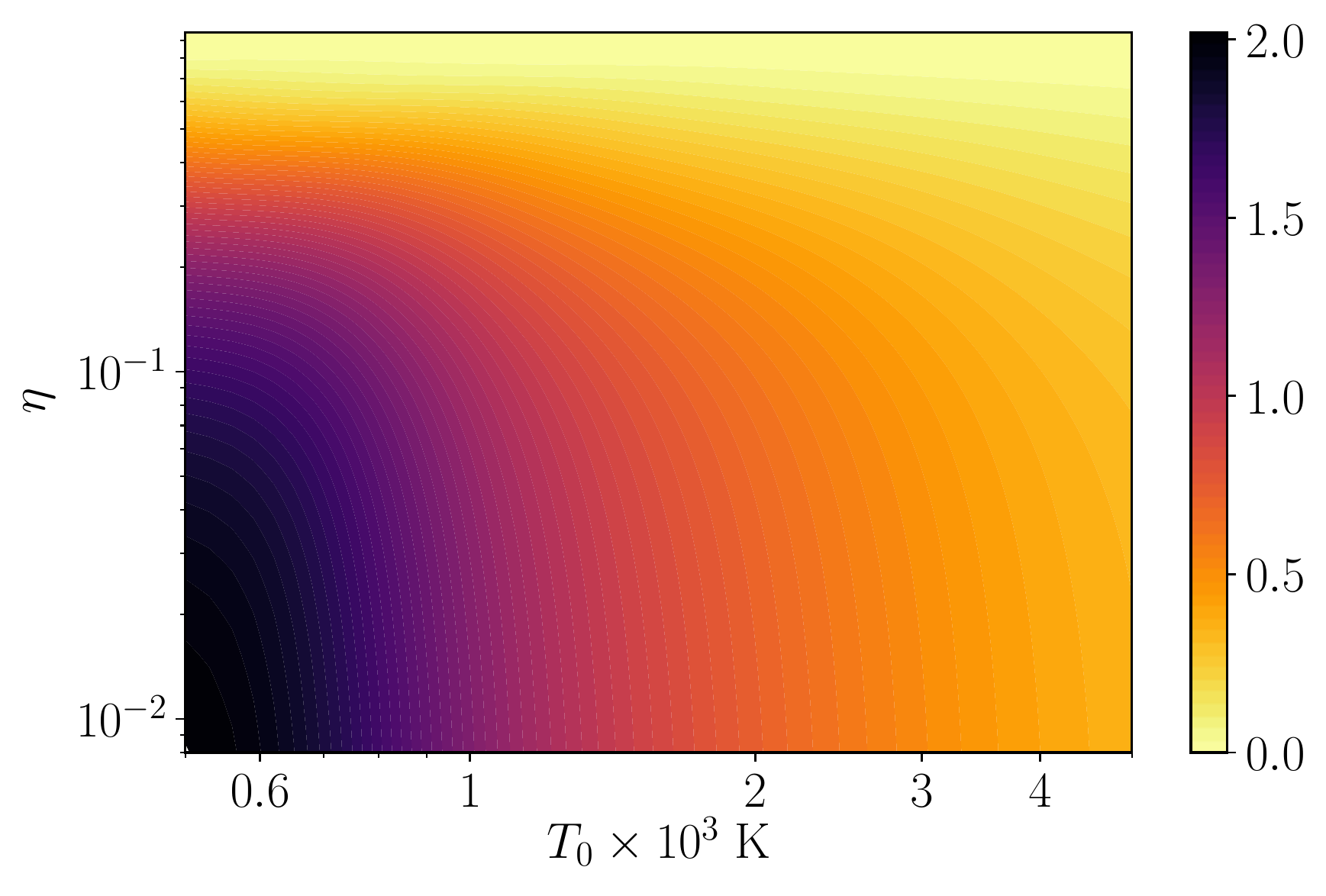}
\caption{Scaled down optical depth $\widehat{\tau} \left( T_0, \eta \right)$ as a function of disk midplane temperature $T_0$ and disk coordinate $\eta$. \label{fig:H2_tau}}
\end{figure}

\begin{figure}
\epsscale{1.2}
\plotone{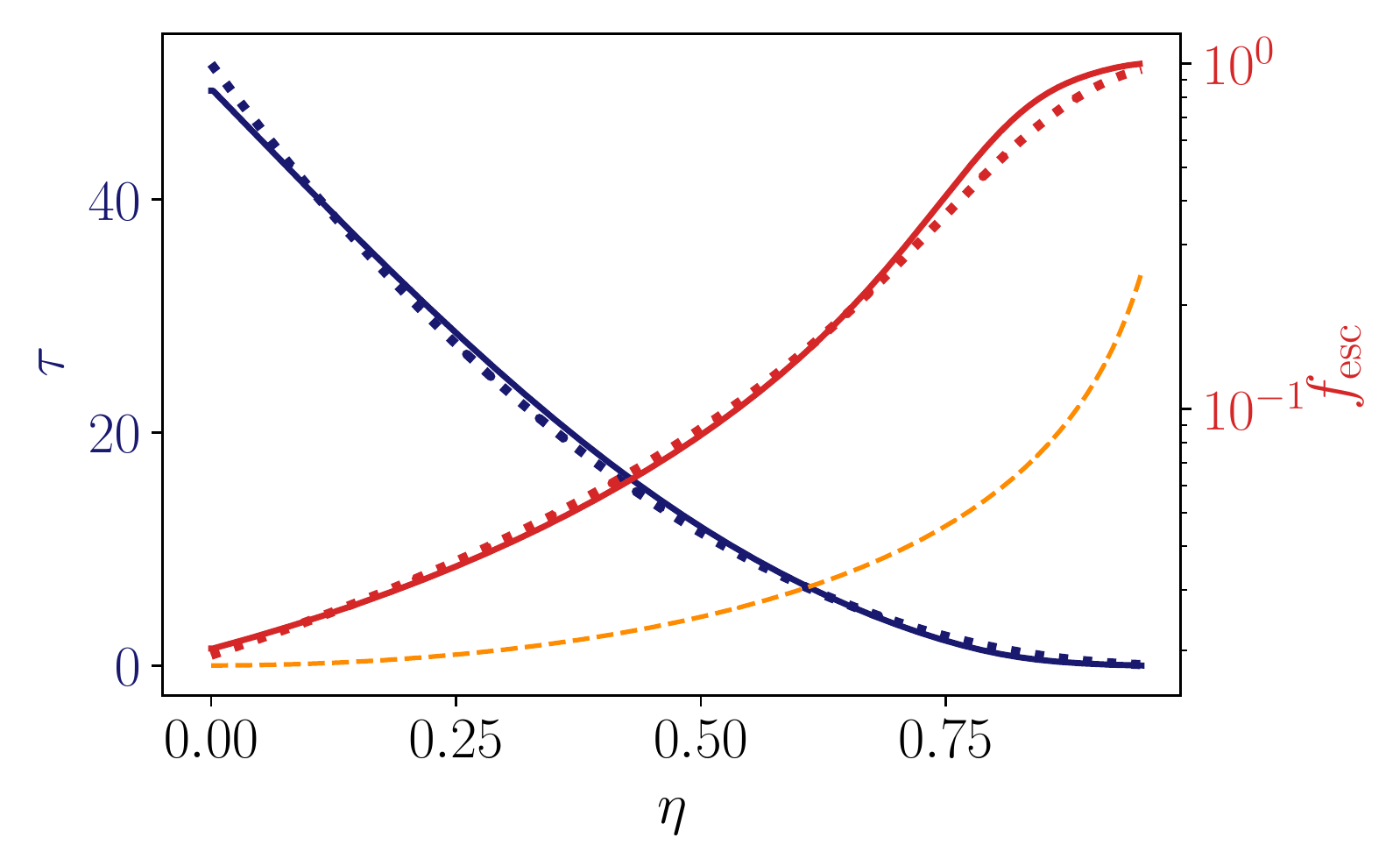} 
\caption{Optical depth (blue) and photon escape fraction (red) for gas at $r=10 {\rm\ AU}$ with $\rho_0= 10^{-10} {\rm\ g\ cm^{-3}}$, $T_0 = 1,500 {\rm\ K}$, and $f_{\rm H_2} = 1$. Solid lines are from numerical integration and the dotted lines use the analytic fitting function from Eq \ref{eq: tau_fit}. As a comparison, the orange dashed line shows $f_{\rm esc}$ computed under the assumption of spherical symmetry.  \label{fig:f_esc}}
\end{figure}

We integrate over Eq (\ref{eq: alpha_formula}) to get $\widehat{\alpha}(T)$. 
For the temperature range in the primordial star forming region, it is sufficient to include the vibrational level v $= 0 - 2$ and rotataional level $J=0-20$ \citep{Yoshida+06}. 
With $\widehat{\alpha}(T)$, $\widehat{\tau} \left( T_0, \eta \right)$ can then be calculated based on Eq (\ref{eq: eta_formula}) and its contour map is shown in Figure \ref{fig:H2_tau}.
$\widehat{\tau} \left( T_0, \eta \right)$ can further be fitted with an analytic expression, 
\begin{eqnarray}
\label{eq: tau_fit}
\widehat{\tau} \left( T_0, \eta \right) = \widehat{\tau}_0 \, \left( \frac{T_0}{10^3 {\rm\ K}} \right)^{c1}   (1-\eta)^{c2} \, .
\end{eqnarray}
The coefficients are listed in Table \ref{tab:tau_fit}. 
With the optical depth, photon escape fraction is then 
\begin{eqnarray}
f_{\rm esc} \equiv \frac{1 - \exp \left( - \tau \right)}{\tau} \, .
\end{eqnarray}
In Figure \ref{fig:f_esc}, we show the optical depth $\tau$ and photon escape fraction $f_{\rm esc}$ at  $r=10 {\rm\ AU}$ in a model with midplane density $\rho_0 = 10^{-10} {\rm g\ cm^{-3}}$ and $T_0 = 1,500 {\rm\ K}$.

\begin{table}
\centering
\begin{tabular}{c|c|c}
$\widehat{\tau}_0$ & c1 & c2 \\
\hline
\hline
$1.34$ & $-0.79$ & $ 2.18$  \\
\end{tabular}
\caption{Coefficients for $\widehat{\tau} \left( T_0, \eta \right) $.} \label{tab:tau_fit}
\end{table}


\section{Fragmentation Conditions}
\label{S: fragmentation}

\subsection{Fiduical Disk Model}
\label{S: fiducial}
Using this approximation for the H$_2$ optical depth, we now seek to characterize conditions for fragmentation in a warm, massive primordial accretion disk.
As accretion flows are less gravitationally stable when $M_d / M_* > 1$, we apply our analysis to early phases, with $M_* = 0.2 \, M_\odot$, and we introduce a fiducial disk model.

Motivated from previous cosmological simulations, we choose surface density $\Sigma$ and midplane temperature $T_0$ as: 
\begin{eqnarray}
\Sigma(r_{10}) &=& \widehat{\Sigma} \; r_{10}^{-1} \, , \\
T_0(r_{10}) &=& \widehat{T}_0 \, r_{10}^{-0.15}  \, ,
\end{eqnarray}
where $r_{10} \equiv r / (10 {\rm\ AU})$, $\widehat{\Sigma} \equiv 0.1 M_\odot / ( \pi \times (10 {\rm\ AU})^2 ) $ and $ \widehat{T}_0 = 1,400 {\rm\ K} $.
The enclosed mass is thus 
\begin{eqnarray}
M_{\rm enc} (r_{10}) = \widehat{M}_{\rm enc} \, \left( r_{10} + 1 \right) \, ,
\end{eqnarray}
with $\widehat{M}_{\rm enc} = 0.2 \, M_\odot$. 
We can now estimate for the disk scale height $H$ and the midplane density $\rho_0$. 
\begin{eqnarray}
H(r_{10})  \equiv \frac{c_s}{\Omega} \approx  \left \{
\begin{tabular}{c l}
$\widehat{H} \; {r_{10}}^{1.425} $ ,  & $r_{10} < 1$ \\
$\widehat{H} \; {r_{10}}^{0.925} $ , & $r_{10} > 1$ 
\end{tabular}
\, , \right .
\end{eqnarray}
\begin{eqnarray}
\label{eq: rho_0}
\rho_0(r_{10})  \equiv \frac{\Sigma}{2H} \approx  \left \{
\begin{tabular}{c l}
$\widehat{\rho}_0 \; {r_{10}}^{-2.425} $ ,  & $r_{10} < 1$ \\
$\widehat{\rho}_0 \; {r_{10}}^{-1.925} $ , & $r_{10} > 1$ 
\end{tabular}
\, , \right . 
\end{eqnarray}
where $\widehat{H} = 6.8 {\rm\ AU}$ and $\widehat{\rho}_0 = 2.8 \times 10^{-11} {\rm\ g\ cm^{-3}}$.
As a result of limited coolants in the early Universe, the temperature in primodial accretion disks typically is at \textit{least} an order of magnitude higher than present-day accretion disks.  The accretion disk is thus supported in part by pressure, with $H/r \sim 0.68$. 

\begin{figure}
\epsscale{1.1}
\plotone{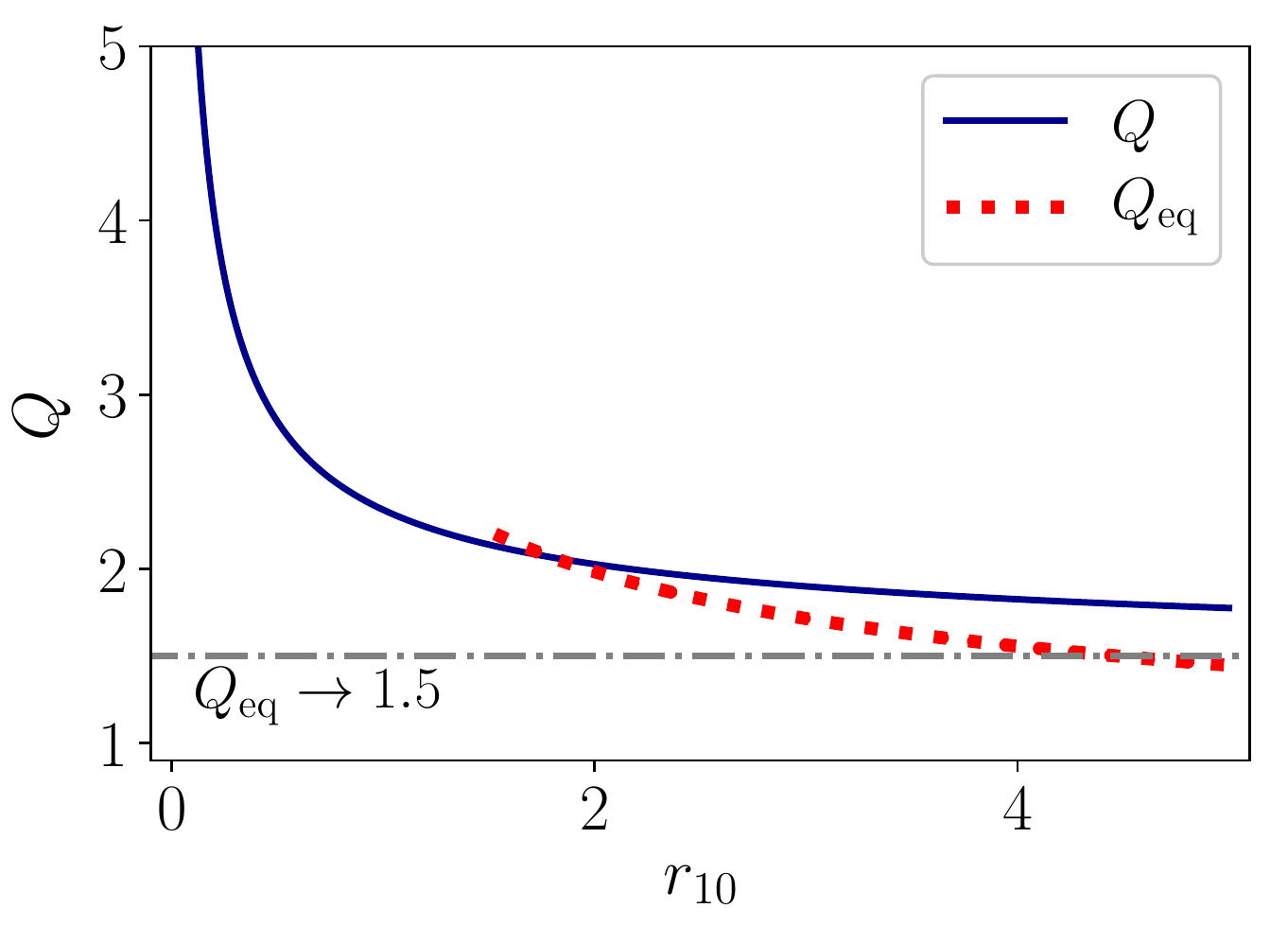} 
\caption{Toomre Q parameter. $Q$ and $Q_{eq}$ are evaluated, respectively, using the fiducial model and under the condition that heating and cooling are balanced. Since $Q$ is much larger than unity, the inner disk is gravitationally-stable against fragmentation. The outer disk (with $Q_{eq} \rightarrow 1.5$) is potentially gravitational unstable to fragmentation under a non-axisymmetric perturbation.  \label{fig:Toomre_Q}}
\end{figure}

The most important diagnostic parameter for fragmentation is the Toomre $Q$ parameter, which is a measure of the pressure and rotational support against self-gravity. The Q parameter is defined as
\begin{eqnarray}
Q \equiv \frac{c_s \kappa }{\pi G \Sigma}  \approx
\left \{
\begin{tabular}{c l}
$\widehat{Q} \; {r_{10}}^{-0.575} $ ,  & $r_{10} < 1$ \\
$\widehat{Q} \; {r_{10}}^{-0.075} $ , & $r_{10} > 1$ 
\end{tabular}
\, , \right . 
\end{eqnarray}
where $\widehat{Q} = 1.9$ and $\kappa$ is the epicycle frequency. Toomre Q is plotted against $r_{10}$ in Fig \ref{fig:Toomre_Q}.
For an axisymmetric perturbation, $Q<Q_{\rm crit} = 1$ is a sufficient condition for fragmentation. 
In addition, for a clump to survive against the background tidal field, the density has to be higher than the Roche density,
\begin{eqnarray}
\rho_{\rm Roche} \approx \frac{M_{\rm enc}}{\pi r^3} \, .
\end{eqnarray}
Since $Q \gg 1$ and $\rho_{\rm Roche } \gtrsim 10^{-8} {\rm\ g\ cm^{-3}}$ in the inner disk with $r \lesssim 3 {\rm\ AU}$, it is unlikely for a clump to fragment due to gravitational instability and survive under the strong tidal field.  
We will also see later that cooling is less efficient in the inner disk, where the disk is likely undergo adiabatic heating. 

\subsection{Cooling and Heating Timescales}
For a gravitationally-unstable cloud to collapse, it must cool down fast enough so that sound speed does not smear out or otherwise disrupt density perturbations. 
Therefore, the general applicable domain is when cooling is both more efficient than net heating, and fast compared to the dynamical time. 

From the first law of thermodynamics, 
\begin{eqnarray}
d \dot{u} = d \dot{q} + \Gamma_{PdV} + d \dot{\mu} \, ,
\end{eqnarray}
where $d \dot{u}$ is the internal energy change rate per volume, $d \dot{q}$ is the non-adiabatic heating/cooling rate per volume, $\Gamma_{PdV}$ is the $PdV$ heating rate per volume, and $d \dot{\mu}$ is the chemical potential changing rate per volume. 
For the purposes of characterizing fragmentation, we note that
\begin{eqnarray}
\Gamma_{PdV} \approx 3 P \, t_{\rm ff}^{-1} \, ,
\end{eqnarray}
where $t_{ff}$ is the free fall time and the chemical potential change is primarily due to the formation and destruction of H$_2$. 
\begin{eqnarray}
d \dot{\mu} = (4.48 {\rm\ eV}) d \dot{n}_{\rm H_2} \,.
\end{eqnarray}
The non-adiabatic processes include viscous heating and radiative cooling:
\begin{eqnarray}
d \dot{q} =  \Gamma_{\rm viscous} - \Lambda_{\rm H_2} - \Lambda_{\rm CIE} .
\end{eqnarray}
$\Gamma_{\rm viscous}$, $\Lambda_{\rm H_2} $, and $\Lambda_{\rm CIE}$ are viscous heating, H$_2$ line cooling and CIE cooling rate. 
For the viscous heating, the macroscopic viscosity due to turbulence is more important than the microscopic viscous effect. It is common to paramatrize turbulence viscosity by an $\alpha$ parameter as $\nu \approx \alpha c_s H$ \citep{Shakura+73}. The viscous heating is thus
\begin{eqnarray}
\Gamma_{\rm viscous }  = \rho \nu \frac{c_s^2}{H^2} = \alpha \rho c_s^2 \Omega \, .
\end{eqnarray}
Since the midplane is usually optically thick to H$_2$ line cooling and optically thin to CIE cooling, we have
\footnote{The cooling rate for optically thin H$_2$ and CIE cooling rates are \citep{Hollenbach+79, Ripamonti+04}:
\begin{eqnarray}
\nonumber
\Lambda_{\rm H_2, thin} / \rho = \frac{\chi_{\rm H} \, f_{\rm H_2}}{m_H} \, 
 \left[  
\mathcal{C}_{\rm H_2} e^{- \left( 0.13/T_3 \right)^3 } +  3 \times 10^{-24} e^{-0.51/T_3} + \right . 
\\ \nonumber
\left. 6.7 \times 10^{-19} e^{- 5.86/T_3 } + 1.6 \times 10^{-18} e^{-11.7/T_3} \right] \, ,
\end{eqnarray}
$$ \Lambda_{\rm CIE, thin} / \rho = 7.2 \times 10^{-2} \, \rho \, T^4 \,  \chi_{\rm H} \, f_{\rm H_2} \, .$$
In the above equations, $\chi_H = 0.76 $ is the hydrogen mass fraction, and $\mathcal{C}_{\rm H_2} = \left( 9.5 \times 10^{-22} T_3^{3.76} \right) /  \left( 1 + 0.12 T_3^{2.1} \right)$.  
}
\begin{eqnarray}
&& \Lambda_{\rm H_2} = f_{\rm esc} \, \Lambda_{\rm H_2, thin} \, , \\
&& \Lambda_{\rm CIE} = \Lambda_{\rm CIE, thin} \, .
\end{eqnarray}
In addition, since chemical potential, $\mu$, acts as a stable agent which tends to reduce the temperature change, it will not by itself lead to fragmentation. We leave the discussion to the next subsection. 

\begin{figure}
\epsscale{1.2}
\plotone{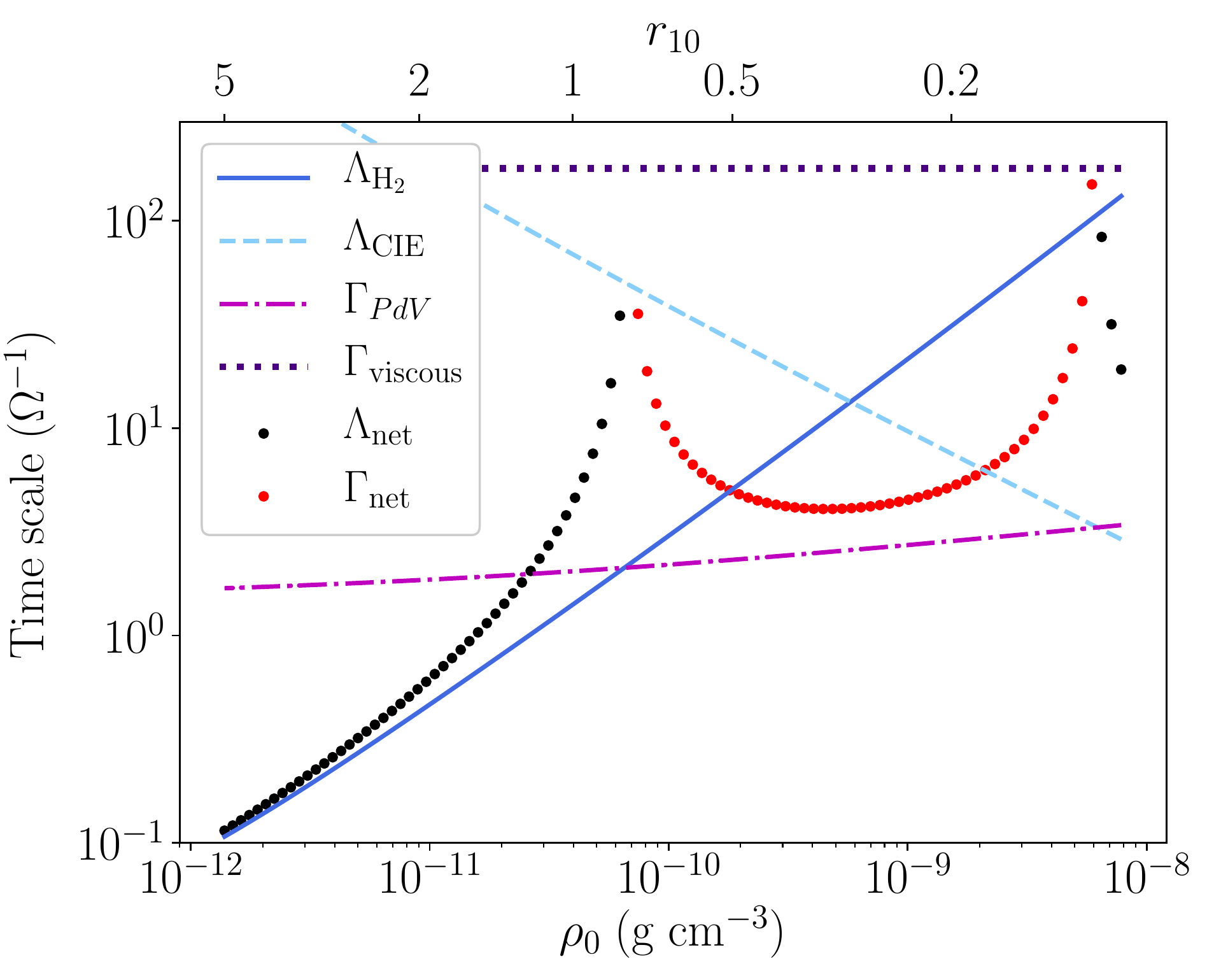} 
\caption{Cooling ($\Lambda$) and heating ($\Gamma$) timescales in the unit of local dynamical time. 
H$_2$ and CIE cooling time are shown in blue solid and dashed lines. 
In the inner disk with $\rho \gtrsim 10^{-9} {\rm\ g\ cm^{-3}}$, cooling is primarily through CIE cooling. H$_2$ cooling becomes more important at the outer disk with low midplane density. The net cooling, shown in black dots, becomes faster than local dynamical time around $r_{10} \gtrsim 1.5$.
$PdV$ and viscous heating are demonstrated in purple dashed-dotted and dotted lines.
Although viscous heating, evaluated at $\alpha = 0.01$, is much slower than most of other processes, $PdV$ heating is the fast when $r_{10} \lesssim 1 $.
As a result, the inner disk with $r_{10} \lesssim 1$ is dominated by heating during the gravitational collapsing phase, which is shown in red dots.  
\label{fig:time_scale}}
\end{figure}

We compare the timescales of each process in Fig \ref{fig:time_scale}. 
In the inner disk, CIE is the dominant channel of cooling. Whereas, H$_2$ cooling takes over CIE cooling at $r \gtrsim 3 {\rm\ AU}$. The net cooling becomes faster than dynamical time at $r \gtrsim 15 {\rm\ AU}$. 
For the heating, it is predominantly through $PdV$ process, which overtakes cooling in the inner disk where $r \lesssim 10 {\rm\ AU}$ with $\rho_0 \gtrsim 10^{-10} {\rm\ g\ cm^{-3}}$. Cooling, and consequently fragmentation, is thus hard to take place in the inner region without reducing the surface density. 

From the energy argument, the outer disk with $r \gtrsim 15{\rm\ AU}$ would have a tendency to fragment. 
However, when the temperature decreases due to cooling, the cooling from ro-vibrational emission and CIE emission also reduces, together with an increased optical depth. One would therefore expect a new equilibrium state where cooling is balanced by heating. 

We provide a heuristic attempt to derive the new equilibrium state. 
The new equilibrium should satisfy $ \Lambda_{\rm cooling} = \Gamma_{\rm heating} $, which leads to
\begin{eqnarray}
f_{\rm esc} \, \Lambda_{\rm H_2, thin} \approx \Gamma_{ PdV} \, .
\end{eqnarray}
Since our interest is in the fragmentation, we focus at $r_{10} > 1$. 
Because $t_{\rm cool} / t_{\rm dyn} < 1$, the density still follows Eq \ref{eq: rho_0} during a cooling time. 
For an optically thick disk with an equilibrium temperature not too far from $10^3 {\rm\ K}$, 
\begin{eqnarray}
f_{\rm esc}  \, \Lambda_{\rm H_2, thin} \approx 
\tau^{-1}  \left( \frac{0.76 \, f_{\rm H_2}}{m_H} \, 9.5 \times 10^{-22} {T_{3, {\rm eq}}}^{3.76} \right) \, .
\end{eqnarray}
Equilibrium temperature is thus
\begin{eqnarray}
T_{\rm eq} = \widehat{T}_{\rm eq}  \,   r_{10}^{-0.55}  
= \left( 1770 {\rm\ K } \right) \,   r_{10}^{-0.55} \, .
\end{eqnarray}
 
Using the $T_{\rm eq}$, we calculate for Toomre parameter at equilibrium state $Q_{\rm eq}$, and show it in Fig \ref{fig:Toomre_Q}. 
\begin{eqnarray}
Q_{\rm eq} = 2.1 \, r_{10}^{-0.275} \, , \quad {\rm\ for\ } r_{10} > 1 \, .
\end{eqnarray}
Since $Q_{\rm eq} \gtrsim 1$, the accretion disk is still stable toward an axisymmetric perturbation through the cooling channel. 
However, for a non-axisymmetric perturbation, \cite{Lau+78} introduce an additional paramter $J \equiv M_{\rm disk} / M_{*}$. 
If $J > 1$, the accretion disk could still be gravitataionally unstable when $Q$ is large but not too far from unity.

\subsection{Chemothermal Instability}
Another possible avenue for fragmentation channel is the chemothermal instability(CTI), which is an extension of thermal instability \citep{Field+65}. 
Thermal instability is known to separate the gas into two phases --- a low-density/high-temperature phase and a high-density/low-temperature phase. 
Once the metal free gas enters a higher density state, the molecular hydrogen fraction increases due to the increasing collision frequency, consequently increasing the CIE cooling rate. As a result, fast cooling is possible and can potentially result in fragmentation \citep{Ripamonti+04, Greif+13}.

To explore the role of CTI, we apply linear analysis following the procedure in \cite{Omukai+03b}. The perturbed energy equation is
\begin{eqnarray}
\label{eq: perturb_energy}
\frac{d \delta u}{dt} = - \delta \mathcal{L} + \varepsilon_{\rm H_2} \delta \mathcal{F}_{\rm H_2} \, ,
\end{eqnarray}
with $u = \frac{3}{2} n_{\rm HI} k_B T + \frac{5}{2} n_{\rm H_2} k_B T $, $\mathcal{L} = f_{\rm esc} \Lambda_{\rm H_2} + \Lambda_{\rm CIE} $, $\varepsilon_{\rm H_2} = 4.48 {\rm\ eV}$, and $\mathcal{F}_{\rm H_2}$ is the molecular hydrogen formation rate. 
Since we are interested in the regime where $\rho \gtrsim 10^{-11} {\rm\ g\ cm^{-3}}$, the rate equation includes only the three body processes. 
\begin{eqnarray}
\mathcal{F}_{H_2} = \frac{d n_{\rm H_2} }{dt} = k_f n_{\rm HI}^3 - k_d n_{\rm H_2} n_{\rm HI} \, ,
\end{eqnarray}
with $k_f$ and $k_d$ being the formation and dissociation rate of molecular hydrogen.
Therefore,
\begin{eqnarray}
\label{eq: perturb_nH2}
\frac{d \delta n_{\rm H_2} }{dt} = \delta \mathcal{F}_{\rm H_2} \, .
\end{eqnarray}
Furthermore, the perturbation is along a constant pressure curve:
\begin{eqnarray}
\label{eq: perturb_P}
\frac{\delta P}{P} = \frac{\delta \rho}{\rho} + \frac{\delta \chi}{\chi} + \frac{\delta T}{T} = 0 \, ,
\end{eqnarray}
where $\chi \equiv 1 -f_{\rm H_2}/2$.

\begin{figure}
\epsscale{1.1}
\plotone{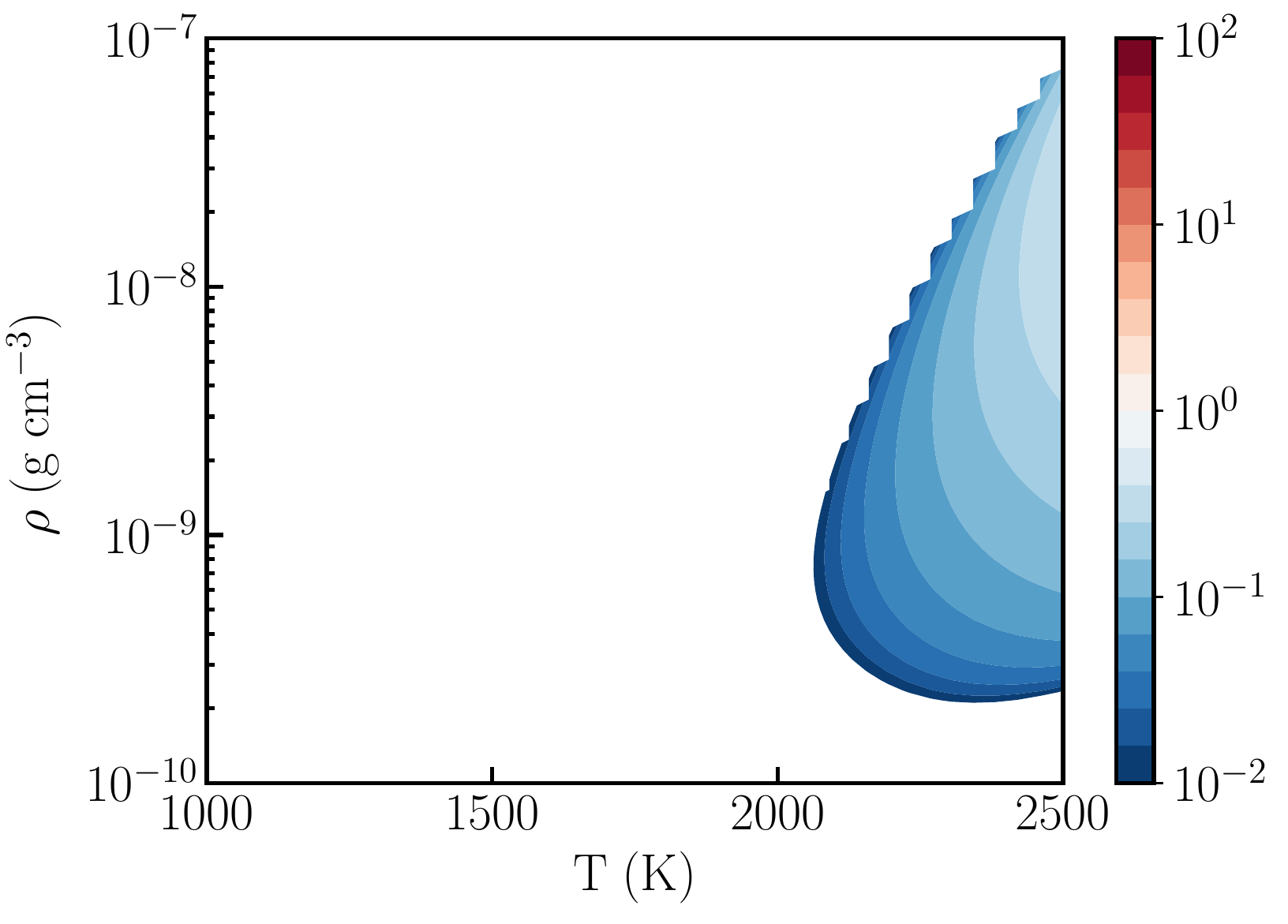} 
\caption{The growth rate, $\gamma_{\rm CTI} \, t_{\rm ff}$, of chemothermal instability. The contour region outlines the parameter space that is unstable to CTI, and $\gamma_{\rm CTI} \, t_{\rm ff}$ is color coded. Since $\gamma_{\rm CTI} \, t_{\rm ff} \ll 1$, CTI is a slow process compared to the free fall time. 
\label{fig:chemo-thermal}}
\end{figure}

Assuming that $\delta \rho, \, \delta T, \, \delta n_{\rm H_2} \propto \exp(\gamma_{\rm CTI} \, t)$, we solve Eq \ref{eq: perturb_energy}, \ref{eq: perturb_nH2} and \ref{eq: perturb_P} together for the growth rate $\gamma_{\rm CTI}$. In addition, the perturbation requires a background equilibrium state, which we derive by setting $d u / dt = 0$. To be specific, we first scan through parameter space $(\rho, T)$ to find a background equilibrium state for $n_{\rm H_2}$. With the equilibrium state, we then solve for three coupled linear equations for $\gamma_{\rm CTI}$. Instability requires $\gamma_{\rm CTI}$ to be real and positive. Fig \ref{fig:chemo-thermal} shows the parameter space that instability takes place. We further color coded $\gamma_{\rm CTI} \, t_{\rm ff}$ and $\gamma_{\rm CTI} \, t_{\rm ff} \gtrsim 1$ indicates a fast growing mode compared to the corresponding free fall time. As shown in Fig \ref{fig:chemo-thermal}, although instability exists when $\rho \gtrsim 10^{-10} {\rm g\ cm^{-3}}$ and $T \gtrsim 2 \times 10^3 {\rm\ K}$, it is slowly growing with $\gamma_{\rm CTI} \, t_{\rm ff} \ll 1$. As a result, CTI is less likely to be an efficient channel that leads to fragmentation.

\section{Survival of Fragmentation}
\label{S: migration}
Although a strong tidal field, coupled with inefficient cooling, inhibits fragmentation in the inner disk, the outer disk is undergoing fast cooling ($t_{\rm cool} \lesssim \Omega ^{-1}$) and could still be gravitationally unstable. What would be the fate of the fragments formed in this region?  Do the fragments migrate inward and merge with the central primary star? 
Could a fragment continue to accrete, and evolve into a massive binary system,
or, would a low mass fragment survive, be scattered away and potentially be found in the local Universe?

Studies of fragmentary migration have largely been confined to studies of planet formation; we apply these analyses here.
We will use the term secondary/clump interchangeably to refer to the fragments, in contrast to using primary/star for the central star.
When the secondary's mass is small and is embedded in the disk, the net torque from the excited density wave pushes the secondary moving inward. This is known as type I migration. 
When the secondary becomes massive enough that its Hill radius exceeds disk scale height, a gap could then be opened up. The migration at this stage is known as type II migration. During this stage, the large inertial of the secondary makes it hard to be further pushed. Type II migration thus corresponds to a slow down of migration. 

\subsection{Type I Migration}
\label{S: type-I}
We first look at the type I migration with a generalized disk model
\begin{eqnarray}
\Sigma_d (r_{10})  = \widehat{\Sigma}_d \, {r_{10}}^{ \sigma - 2 } \, ,
\end{eqnarray}
where $\Sigma_d$ is disk surface density. We use $\Sigma _d$ to distinguish it from $\Sigma$ in the fiducial model in S \ref{S: fiducial}.
Allowing the total disk mass enclosed within $50 {\rm\ AU}$ to be a fraction, $f_d$, of the star mass, we have
\begin{eqnarray}
\widehat{\Sigma}_d = f_d \, m_* \sigma ' \, \widehat{\Sigma}_{d, 0} \, ,
\end{eqnarray}
with $\widehat{\Sigma}_{d, 0} \equiv M_{\odot} / 2 \pi ( 10{\rm\ AU} ) ^2 $, $M_* \equiv m_* \, M_\odot$, 
and $\sigma' \equiv \sigma / \left( 5^{\sigma} - 1 \right)$.
Under a thin disk and isothermal assumption, \cite{Tanaka+02} provides a migration formula that accounts for the torque from both Lindbald and corotation radius. 
The migration time
\begin{eqnarray}
\label{eq: t_m}
t_{\rm m} &=& \left( 4.9 - 1.1 \sigma \right)^{-1} 
\left( \frac{M_s}{M_*} \right)^{-1}  \left( \frac{M_*}{\Sigma_d \, r_s^2} \right) h_s^2 \Omega_s^{-1} \, , \\ 
&=&
\label{eq: t_m1}
2 \, f_d^{-1} r_{s, 10}^{3/2} \, T_{s, 3} \, m_*^{-3/2}   \left( \frac{m_*}{m_s} \right) {\rm\ yr} \, ,
\end{eqnarray}
where the subscript $s$ denotes the disk condition at the location of the secondary, and the last equality uses $\sigma = 1$. 
As shown in Eq \ref{eq: t_m1}, a long migration time requires a low disk mass and a low clump mass, a large separation from the primary, and a low primary mass.

For a secondary to survive, the migration must be slower than the primary's Kelvin-Helmholtz time, where the UV photon from the central star photo-evaporates the accretion disk. 
\cite{McKee+08} estimate the photo-evaporation rate in a primordial protostellar disk:
\begin{eqnarray}
\dot{M}_{\rm evap} = 5.4  \times 10^{-7}  \left( 1 +f_d \right)^{1/2} m_*^{5/4} \, M_\odot {\rm\ yr^{-1}}  \, ,
\end{eqnarray}
where we have assumed the ionized gas due to photoionization heating has a temperature around $2.5 \times 10^4 {\rm\ K}$.  
The photoevaporation time $t_{\rm evap}$ is thus
\begin{eqnarray}
t_{\rm evap} &=& \frac{M_d}{\dot{M}_{\rm evap}} 
= 1.9 \times 10^6  \frac{f_d}{\sqrt{1+f_d}} \, m_*^{-1/4} {\rm\ yr} \, .
\end{eqnarray}

\begin{figure}
\epsscale{1.1}
\plotone{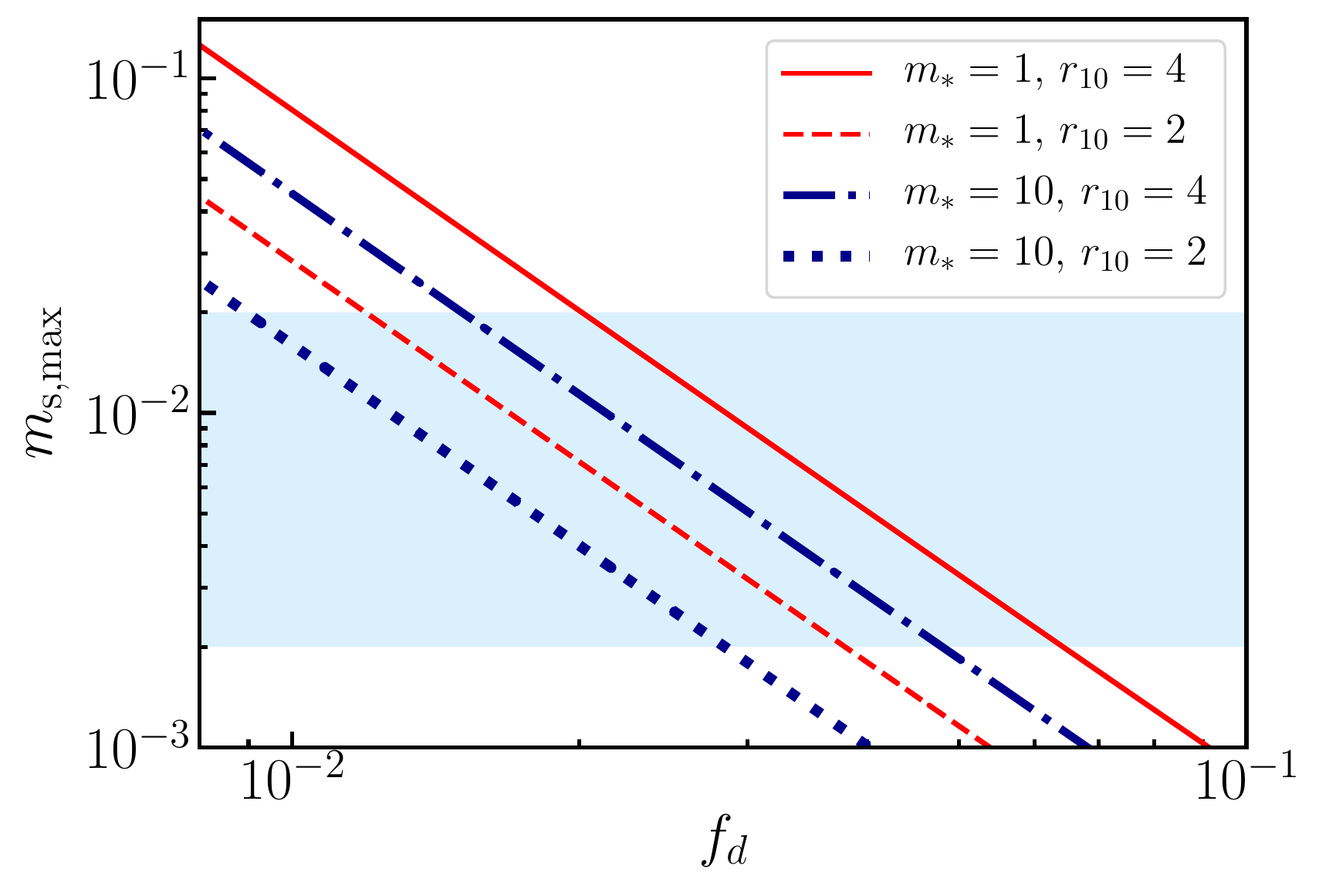} 
\caption{Critical condition for type I migration. Given the central stellar mass and the location of a clump, the plot shows the maximum allowd clump mass that will survive under type I migration. The blue shaded region indicates the Jeans mass of $\rho = 10^{-10} - 10^{-8} {\rm\ g\ cm^{-3}}$. \label{fig:mig_1}}
\end{figure}

Comparing the timescale between migration time and photo-evaporation time, we derive the the clump surviving condition 
\begin{eqnarray}
\frac{t_m}{t_{\rm evap}} = 10^{-6} f_d^{-2} r_{s, 10}^{3/2} T_{s, 3} m_*^{-5/4} \left( \frac{m_*}{m_s} \right) > 1 \, .
\end{eqnarray}
The critical condition for maximum clump mass is then
\begin{eqnarray}
m_s < 10^{-6} f_d^{-2} \, r_{s, 10}^{3/2} \, T_{s, 3} \,  m_*^{-1/4} \, .
\end{eqnarray}
We show the critical condition for a few choices of primary mass and separation in Fig \ref{fig:mig_1}, where we also set $T_{s, 3} = 1$. We focus on the mass range $m_s = 2 \times 10^{-3} - 2 \times 10^{-2}$, which corresponds to Jeans mass with $\rho = 10^{-10} - 10^{-8} {\rm\ g\ cm^{-3}}$. Even for the most compact clump, it requires $f_d < 0.1$ for it to survive against type I migration. Effectively, most of the clumps would migration inward fast through type I migration. 

\subsection{Type II Migration}
Presumably, the secondary is massive enough to carve out a gap around it. Its Hill radius should be comparable to disk scale height. It leads to
\begin{eqnarray}
\label{eq: Hill}
\frac{M_s}{M_*} \gtrsim 3 \, h^3 \, .
\end{eqnarray}
If we approximate the disk mass around the secondary to be $M_{\rm d, s} \sim \pi \Sigma_d r_H^2$, with $r_H$ represents the Hill radius, one can show that $M_s > M_{\rm d, s}$ is almost always true. 
As the inertia of the secondary is much higher, the clump becomes harder to move. As a result, gap opening corresponds to a slow down of the migration. We therefore treat the location of gap opening as the final location of the secondary. The density of a clump would also need to be higher than Roche density in order to survive.

\begin{figure}
\epsscale{1.1}
\plotone{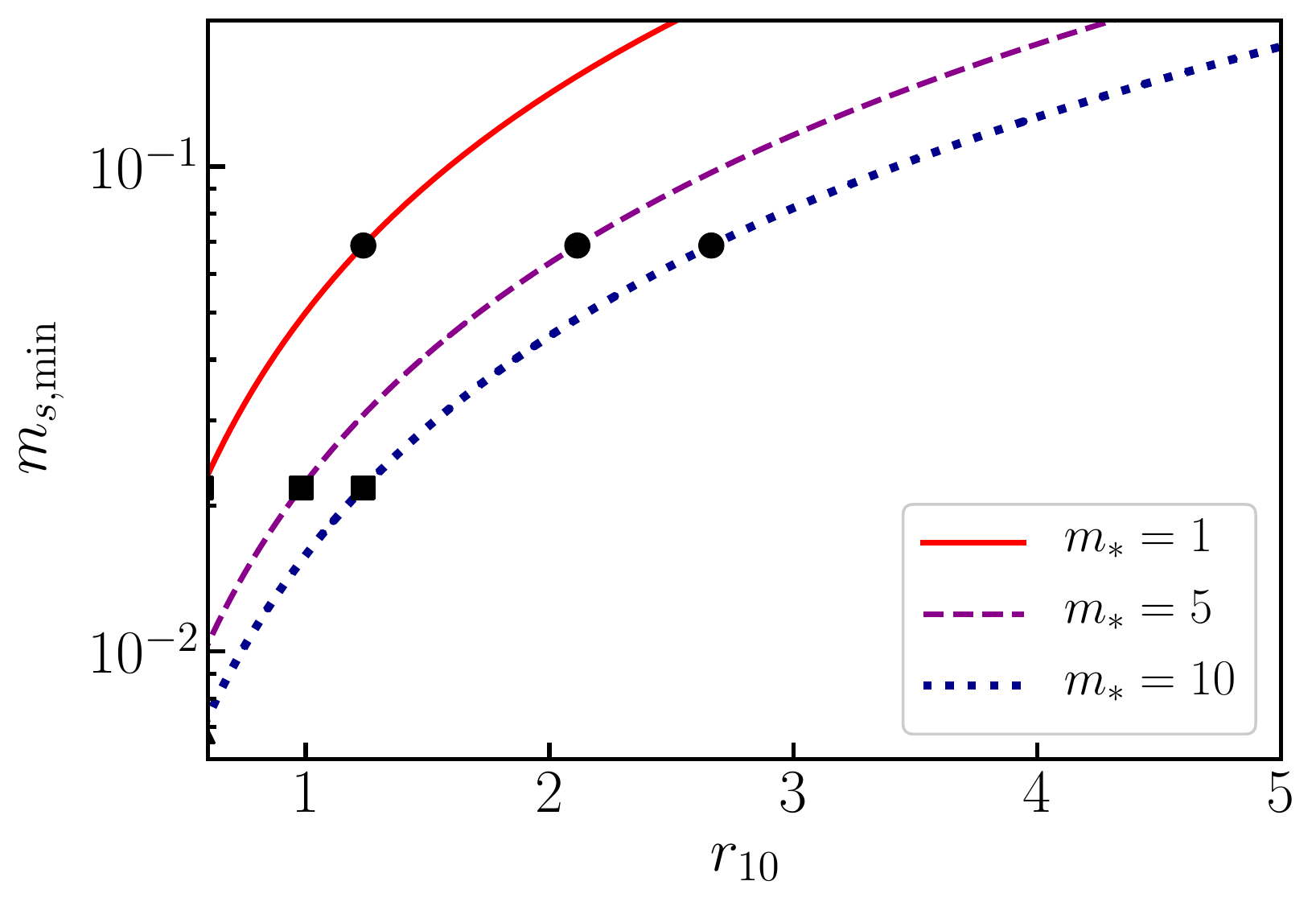} 
\caption{Minimum required clump mass for gap opening. The circles and squares represent the Roche limit for clump with density $\rho = 10^{-10} {\rm\ and\ } 10^{-9} {\rm\ g\ cm^{-3}}$ at the corresponding central stellar mass.  \label{fig:mig_2}}
\end{figure}

Fig \ref{fig:mig_2} shows the required minimum clump mass to open up a gap. The circles, squares and triangles represent the Roche limit for clump with density $\rho = 10^{-10}, \, 10^{-9}, \, 10^{-8} {\rm\ g\ cm^{-3}}$ at the corresponding central stellar mass. 
For a clump mass above $m_s \sim 0.1$, a gap would be opened up and migration would slow down, if the clump density becomes denser than $10^{-10} {\rm\ g\ cm^{-3}}$. We then expect a zone between $20 - 40 {\rm\ AU}$ that traps most of clumps with $m_s > 0.1$. Consequently, a clump would likely run into another one in this region and gradually merge --- leading to a much more massive final clump. Numerical simulations have also found that clump-clump merger is the primary channel for its growth, e.g. \cite{Greif+12}, \cite{Stacy+16}, \cite{Susa+19}.
Also note that, for clumps with lower mass, the gap opens within $20 {\rm\ AU}$. From the Roche limit argument, the clump needs to be denser than $10^{-9} {\rm\ g\ cm^{-3}}$ in order to survive against the background tidal field.

After the clump formation, the gravitational interaction between clumps would have a significantly impact on the subsequent evolution of accretion flow. 
The spiral arm due to the presence of a clump could accelerate the angular momentum transport through gravitational torque; The overdensity in the spiral arm favors fragmentation; The presence of a clump also introduces vortical fluid structure around it, which then affects the subsequent fragmentation.

\section{Clump Surviving Condition}
\label{S: clump}
As demonstrated in Fig \ref{fig:mig_1}, a clump migrates inward fast through type I migration. A plausible way for the clump to survive against migration is through gap opening \citep{Chon+19}. 
Gap opening requires either a massive clump or a massive central star with $M_* \gtrsim 10 M_\odot$ (see Fig \ref{fig:mig_2}). In this section, we look at these two possible mechanisms for gap opening.

\subsection{Gap Opening through a Massive Clump}
For a clump to grow in mass, it must be able to radiate away the excess gravitational potential energy and cool down. The cooling process could be a combination of both convective cooling and radiative cooling. 

We first consider the possibility of having convection. Convection happens when the temperature gradient is steeper than the adiabatic temperature gradient:
\begin{eqnarray}
\vert \frac{dT}{dR} \vert > \vert \frac{dT}{dR} \vert_{\rm ad}  \, .
\end{eqnarray} 
We look at a model where a spherical region in a uniform medium collapses into a uniform clump. From the energy argument, part of the gravitational potential energy would be thermalized and participate into thermal and chemical binding energy:
\begin{eqnarray}
\label{eq: clump_thermal}
f_\Phi \Delta \Phi_g = \Delta U_s + \varepsilon_{\rm H_2} \Delta N_{\rm H_2} \, .
\end{eqnarray}
Note that it is important to include chemical binding energy in the energy equation, since an increased temperature would favor a dissociation process which then extracts energy from the thermal reservoir and compensates for the temperature change. 
In the above equation, $f_\Phi$ is the fraction of gravitational potential energy that has been thermalized and we set it to $0.5$ throughout this work. $\Phi_g \equiv G M_s^2 / R_s$ is the gravitational potential energy and $R_s$ is the clump radius. $U_s$ is the total internal energy and $\Delta N_{\rm H_2}$ is the change of total molecular hydrogen number, which is solved assuming chemical equilibrium. 
In addition, we ask the final clump to be in force equilibrium, i.e., 
\begin{eqnarray}
\label{eq: clump_virial}
P_{s, f} = P_{\rm ext} + \frac{G M_s}{R_{s, f}} \rho_f \, ,
\end{eqnarray}
where $P_s$ and $P_{\rm ext}$ are the pressure of the clump and the pressure from the external medium. The subscript $f$ represents the final states after gravitational collapse. 
Solving Eq \ref{eq: clump_thermal} and \ref{eq: clump_virial} together, we can estimate for $\vert dT / dR \vert $.

\begin{figure}
\epsscale{1.1}
\plotone{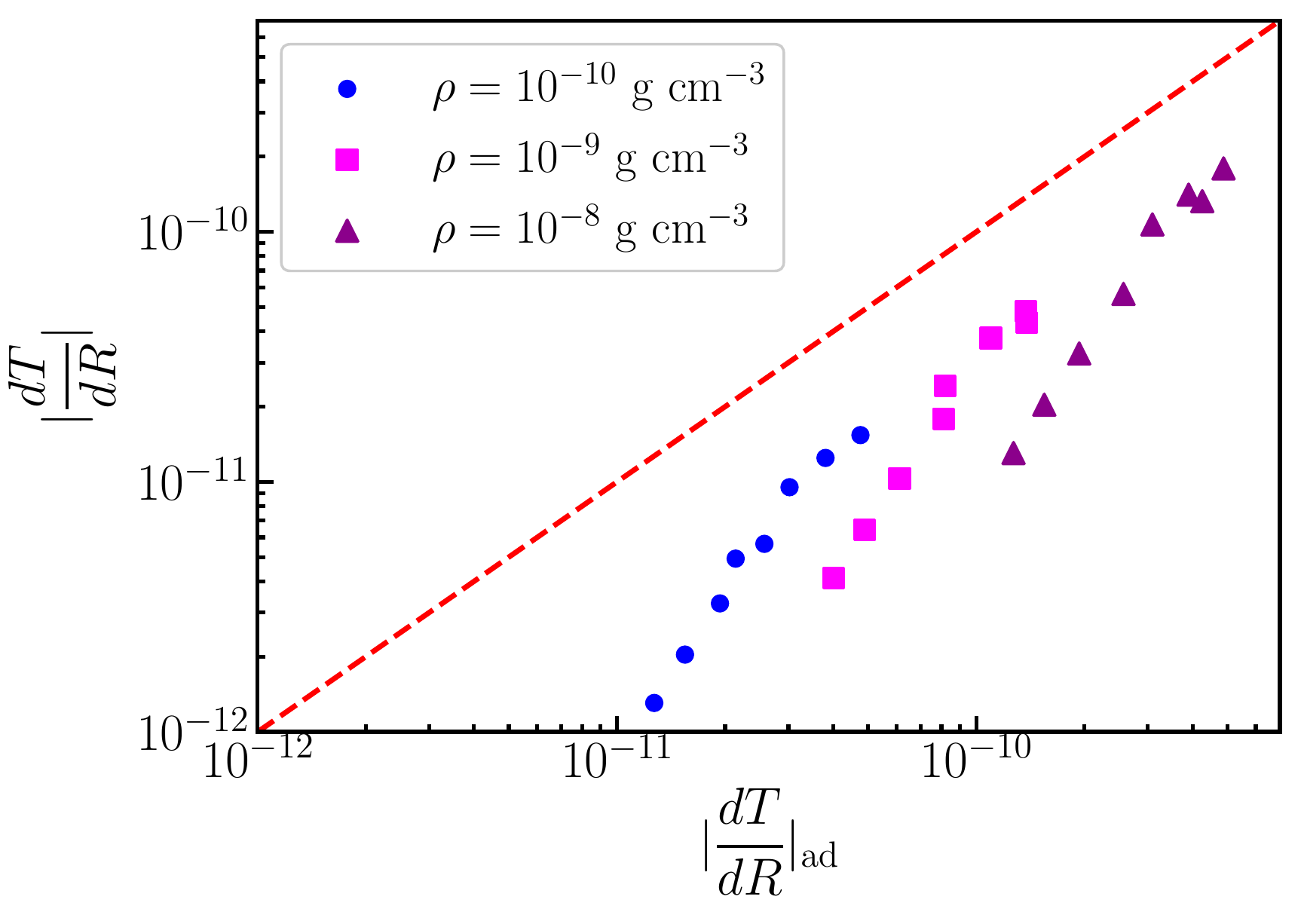} 
\caption{Convection condition for a clump. The red dashed line indicates the limit above which convection takes place. Since all clumps that we considered are below the convection limit, convection cooling would not be an efficient cooling channel.  \label{fig:clump_convection}}
\end{figure}

We look at the collapse from a medium with densities, $\rho=10^{-10} , \, 10^{-9} , \, 10^{-8} {\rm\ g\ cm^{-3}}$, and $T=10^3 {\rm\ K}$ in all cases. The initial gravitational unstable radius increases from half of the Jeans wavelength all the way until the final density becomes less than one and a half of the initial density. In figure \ref{fig:clump_convection}, we plot $\vert dT / dR \vert$ against $\vert dT / dR \vert_{\rm ad} $. The convection condition, $\vert dT / dR \vert = \vert dT / dR \vert_{\rm ad}$, is then shown in red dashed line, below which a clump is stable against convection. Since all the clump satisfy $\vert dT / dR \vert < \vert dT / dR \vert_{\rm ad}$, convection is unlikely to be an efficient cooling channel in a fragmenting clump. 

\begin{figure}
\epsscale{1.2}
\plotone{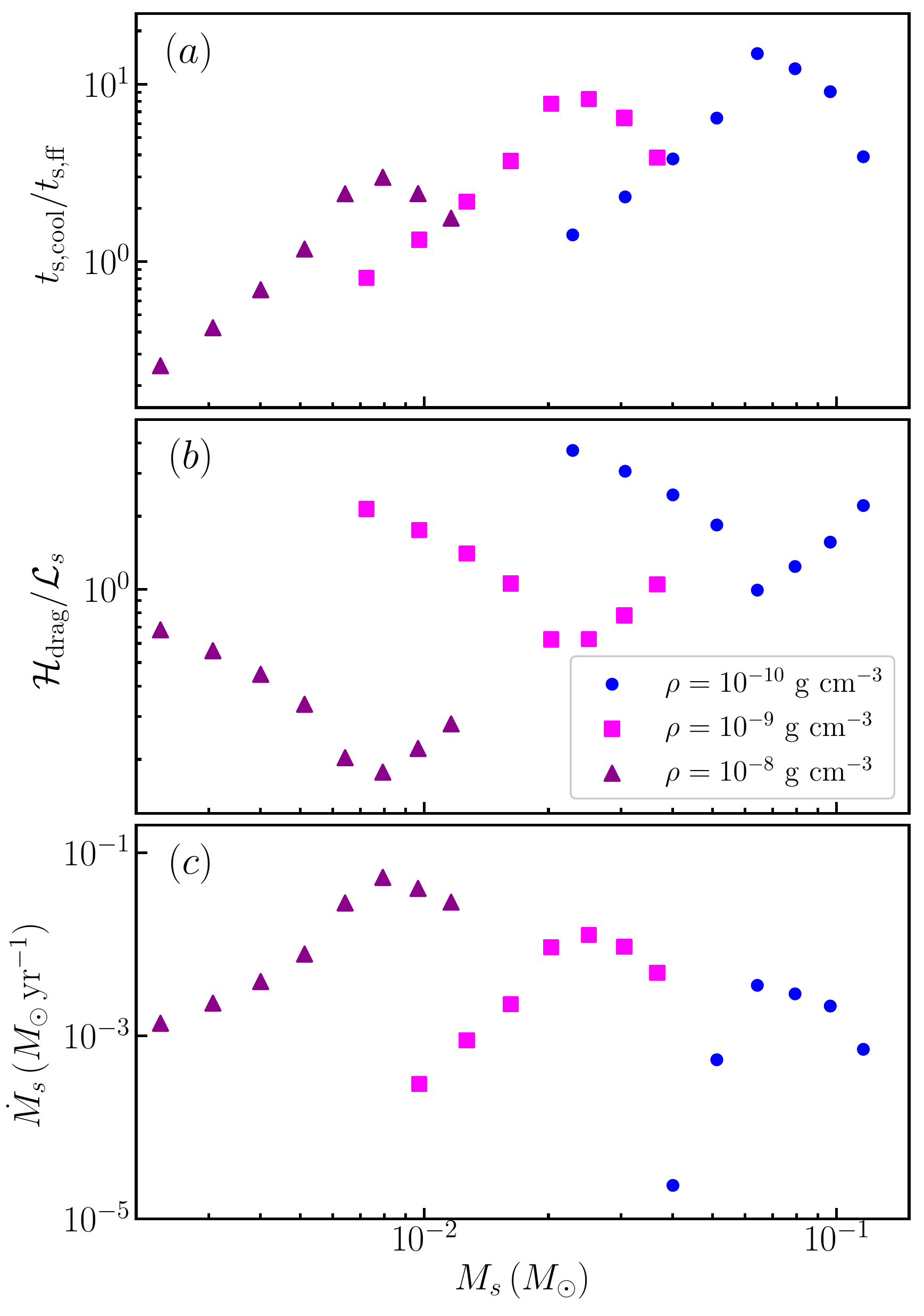} 
\caption{Cooling and accretion processes associated with a clump. Panel (a) shows the cooling time, in the unit of free fall time, for a clump to radiate away half of its gravitation potential energy during gravitataional collapse. Panel (b) shows the ratio between aerodynamic heating, $\mathcal{H}_{\rm drag}$, and radiative cooling, $\mathcal{L}_{s}$ during inward migration. The clump accretion rate is shown in panel (c), with an assumption that it is able to fully radiate away half of its gravitational potential energy and become compact.   \label{fig:clump_all}}
\end{figure}

We then look at the cooling efficiency through radiative cooling. Assuming the other half, $1-f_\Phi$, of the gravitational energy is radiated away, the cooling time is 
\begin{eqnarray}
t_{\rm s, cool} = \frac{ (1-f_\Phi) \Delta  \Phi_g }{ \mathcal{L}_s } \, ,
\end{eqnarray}
with $\mathcal{L}_s \approx \mathcal{L}_{\rm s, rad}$ being the clump luminosity. As shown in Fig \ref{fig:clump_all}(a), a dense clump with sufficient CIE cooling is able to radiate away the gravitational potential energy. Whereas, a clump collapse from a relative low density gas contains more gravitational potential energy ($\propto \rho^{-3/2}$), but has a lower cooling rate per volume. As a result, cooling becomes slower than a free fall time.

In the end, we consider the possibility that a clump could cool while it migrates inward since $t_{\rm s, cool}/ t_{\rm mig} < 1$. In order for a clump to cool during migration, cooling needs to be more efficient than heating caused by aerodynamic drag. The aerodynamic heating is 
\begin{eqnarray}
\mathcal{H}_{\rm drag} = F_d \times v_{\rm diff} \, ,
\end{eqnarray}
where $F_d$ is the aerodynamic drag force on the clump and $v_{\rm diff}$ is the velocity difference between a Keplerian rotating clump and the partially pressure supported accretion flow. For a sphere, 
\begin{eqnarray}
F_d = \frac{1}{2} C_D \rho_{\rm ext} v_{\rm diff}^3 \times \left( \pi R_s^2 \right) \, ,
\end{eqnarray}
with $C_D \sim 0.5$ in a turbulent flow, and $\rho_{\rm ext}$ is the density of the surrounding medium. And, we have $v_{\rm diff} = f_v \times v_{\rm Kep}$, where we choose $f_v = 0.5$ and $v_{\rm Kep}$ is the local Keplerian velocity. 
Consider a clump at $r \sim 40 {\rm\ AU}$, we compare the aerodynamic heating and radiative cooling in Fig \ref{fig:clump_all}(b). 
Again, for a dense clump with $ \rho_s \gtrsim 10^{-8} {\rm\ g\ cm^{-3}}$, increased CIE cooling would allow cooling and further accretion. For a lower density clump, $\mathcal{H}_{\rm drag} / \mathcal{L}_s \sim \mathcal{O}(1)$. As a result, cooling during migration would not be an efficient channel, either.  Therefore, a massive clump with $m_s > 0.01$ needs to first form a compact, high-density core, through which the clump could then gradually grow in mass and simultaneously radiate away any excess gravitational potential energy. 
Note that this is an optimistic estimation, since we do not consider the condition where the clump is shock heated by encountering a propagating density wave.

Furthermore, once a clump manages to form a dense core, clump accretion rate scales with the net cooling rate, 
\begin{eqnarray}
\label{eq: clump_accrete}
\dot{M}_s \sim \frac{ \mathcal{L} _{s, f}- \mathcal{H}_{\rm drag} }{ (1 - f_\Phi)  G M_s / R_s } \, .
\end{eqnarray}
As shown in Fig \ref{fig:clump_all}(c), the accretion rate could reach $\dot{M}_s \gtrsim 10^{-3} \, M_\odot {\rm yr^{-1}}$. 
This fast growing clump would be able to open up a gap, and become a long-lived fragment. However, the mass growth will ultimately be impeded by gap opening. By the time of gap opening, the accretion rate would be limited by the amount of gas available, instead of the cooling efficiency discussed in Eq \ref{eq: clump_accrete}. 
On the other hand, for a clump with $\rho \lesssim 10^{-9} {\rm\ g\ cm^{-3}}$, the clump accretion rate is limited by the cooling efficiency due to the lack of sufficient CIE cooling. Although their initial mass is heavier due to a larger Jeans length, its mass most likely stays in the sub-solar range for the majority of time during its inward migration.

\subsection{Gap Opening through a Massive Central Star}
For a clump without a dense core, and thus lack of efficient cooling, we further look at the possibility of survival in the case of a central, fast-growing star. 
As demonstrated in Fig \ref{fig:mig_2}, gap opening is more likely to happen when the central star becomes massive and provides a stronger gravity that tones down $H/r$. Once the Hill sphere becomes comparable to disk scale height, the excited density wave pushes away the surrounding flow. Surface density decreases. As a result, the photon escape probability increases and radiative cooling becomes efficient. 

In this scenario, a fast accretion rate with $t_{\rm accrete} / t_m < 1$ is needed. We then have 
\begin{eqnarray}
\label{eq: clump_fast_accrete}
\frac{ \dot{M}_* }{M_\odot {\rm\ yr^{-1}}} \gtrsim 2 \times 10^{-3} 
\left( \frac{f_d}{0.1} \right) \left( \frac{r_{10}}{4} \right)^{-\frac{3}{2}} 
\left( \frac{m_*}{10} \right)^{\frac{3}{2}} \left( \frac{m_s}{0.01} \right) \, .
\end{eqnarray}
Since all the brackets on the right-hand side of Eq \ref{eq: clump_fast_accrete} are at the order of unity, it then requires $\dot{M}_* \gtrsim 2 \times 10^{-3} M_\odot {\rm\ yr^{-1}} $. A gap opening process through a fast growing central star is thus plausible, and preferentially occurs in cases with a rapidly-accreting central star.

For a clump that forms initially at the outer disk, the condition might be close to $\rho = 10^{-10} - 10^{-9} {\rm\ g\ cm^{-3}}$. Since cooling is initially slow, the advantage of being massive allows it to carve out a gap as long as the central star becomes massive during its inward migration. Once the surface density decreases, cooling starts to pick up. A dense central core gradually emerges, which then enables further clump accretion.

\section{Conclusion}
\label{S: conclusion}

We provide an analytic expression for H$_2$ ro-vibrational line cooling optical depth in a metal-free accretion disk.
Using this expression, we study relevant fragmentation conditions during the central star's infant phase, with $M_* = 0.2 \, M_\odot$ and $M_d / M_* > 1$.
We found that, in the inner disk with $r \lesssim 10 {\rm\ AU}$, H$_2$ cooling is attenuated significantly and that $PdV$ heating overtakes the cooling; the accretion flow is unlikely to substantially cool without reducing the disk's surface density. 
We also look at the possibility of fragmentation through chemothermal instability, and find that it is typically slower than the relevant timescales, with a grow rate $\gamma_{\rm CTI} \, t_{\rm ff} \ll 1$. Under our analytic expressions, fragmentation is unfavorable (but not excluded) in the inner disk; harmonizing this result with simulation results that \textit{demonstrate} inner-disk fragmentation will be pursued in a subsequent paper.

In contrast, in the outer disk, cooling starts to become dynamical with $t_{\rm cool} \lesssim t_{\rm dyn}$. 
By balancing cooling with $PdV$ heating, we obtain a Toomre Q approaching to $1.5$ when $r \gtrsim 40 {\rm\ AU}$. 
The accretion flow could potentially be unstable to non-axisymmetric perturbation, forming a collapsing clump. 
In our analysis, we demonstrate that majority of fragmentation should be initially formed within the outer disk. 

The long-term survival of these clumps requires evaluating survival conditions for Type I and Type II migration.  Type I migration is effectively fast that a clump would merge with the central star as long as $f_d \gtrsim 0.1$. However, a clump would survive against migration once a gap is carved out. Two possible mechanisms for gap opening are studied: through massive clumps and through a fast-growing central star.

In order for a clump to become massive, it needs to radiate away the gravitataional potential energy released from the clump accretion process. Since convection is barely operating within a clump, cooling is primarily through H$_2$ radiative cooling. In addition, for a clump to accrete mass during inward migration, radiative cooling must be more efficient than aerodynamic heating. As a result, a clump could only become massive through an existence of a dense core with $\rho \gtrsim 10^{-8} {\rm\ g\ cm^{-3}}$ that provides a compelling CIE cooling. In addition, for a clump with $\rho \lesssim 10^{-9} {\rm\ g\ cm^{-3}}$, the accretion rate is limited by cooling efficiency and would be slowly growing with clump accretion rate less than $10^{-3} M_\odot {\rm\ yr^{-1}}$.

Another gap opening mechanism is through a fast growing central star. Once the central star becomes massive, its gravity dominates the star-disk system and reduces disk scale height. Consequently, a gap would then be opened up as long as $m_s \gtrsim 0.01$.
In this scenario, a dense central core is not necessary, but the central star is required to have a fast accretion rate $\dot{M} \gtrsim 2 \times 10^{-3} M_\odot {\rm\ yr^{-1}}$. Through this process, a clump gradually collapses due to the decreased surface density, and proceeds to further accretion.

\acknowledgments
W.-T. L. and M.J.T. are supported in part by the Gordon and Betty Moore Foundation’s Data-Driven Discovery Initiative through Grant GBMF4561 to Matthew Turk. W.-T. L. acknowledges government scholarship to study aboard from the ministry of education of Taiwan, and a support from the National Science Foundation under Grant No. PHY-1430152 (JINA Center for the Evolution of the Elements).


\begin{thebibliography}{}
\expandafter\ifx\csname natexlab\endcsname\relax\def\natexlab#1{#1}\fi

\end{thebibliography}


\begin{thebibliography}{}

\bibitem[{{Abel} {et~al.}(2002){Abel}, {Bryan}, \& {Norman}}]{Abel+02}
{Abel}, T., {Bryan}, G.~L., \& {Norman}, M.~L. 2002, Science, 295, 93

\bibitem[{{Aoki} {et~al.}(2014){Aoki}, {Tominaga}, {Beers}, {Honda}, \&
  {Lee}}]{Aoki+14}
{Aoki}, W., {Tominaga}, N., {Beers}, T.~C., {Honda}, S., \& {Lee}, Y.~S. 2014,
  Science, 345, 912

\bibitem[{{Barkana} \& {Loeb}(2001)}]{Barkana+01}
{Barkana}, R., \& {Loeb}, A. 2001, \physrep, 349, 125

\bibitem[{{Bowman} {et~al.}(2018){Bowman}, {Rogers}, {Monsalve}, {Mozdzen}, \&
  {Mahesh}}]{Bowman+18}
{Bowman}, J.~D., {Rogers}, A.~E.~E., {Monsalve}, R.~A., {Mozdzen}, T.~J., \&
  {Mahesh}, N. 2018, \nat, 555, 67

\bibitem[{{Bromm} {et~al.}(2002){Bromm}, {Coppi}, \& {Larson}}]{Bromm+02}
{Bromm}, V., {Coppi}, P.~S., \& {Larson}, R.~B. 2002, \apj, 564, 23

\bibitem[{{Chon} \& {Hosokawa}(2019)}]{Chon+19}
{Chon}, S., \& {Hosokawa}, T. 2019, \mnras, 488, 2658

\bibitem[{{Clark} {et~al.}(2011){Clark}, {Glover}, {Smith}, {Greif}, {Klessen},
  \& {Bromm}}]{Clark+11}
{Clark}, P.~C., {Glover}, S.~C.~O., {Smith}, R.~J., {et~al.} 2011, Science,
  331, 1040

\bibitem[{{Ewall-Wice} {et~al.}(2018){Ewall-Wice}, {Chang}, {Lazio},
  {Dor{\'e}}, {Seiffert}, \& {Monsalve}}]{Ewall-Wice+18}
{Ewall-Wice}, A., {Chang}, T.-C., {Lazio}, J., {et~al.} 2018, \apj, 868, 63

\bibitem[{{Feng} \& {Holder}(2018)}]{Feng+18}
{Feng}, C., \& {Holder}, G. 2018, \apjl, 858, L17

\bibitem[{{Field}(1965)}]{Field+65}
{Field}, G.~B. 1965, \apj, 142, 531

\bibitem[{{Greif}(2014)}]{Greif+14}
{Greif}, T.~H. 2014, \mnras, 444, 1566

\bibitem[{{Greif} {et~al.}(2012){Greif}, {Bromm}, {Clark}, {Glover}, {Smith},
  {Klessen}, {Yoshida}, \& {Springel}}]{Greif+12}
{Greif}, T.~H., {Bromm}, V., {Clark}, P.~C., {et~al.} 2012, \mnras, 424, 399

\bibitem[{{Greif} {et~al.}(2013){Greif}, {Springel}, \& {Bromm}}]{Greif+13}
{Greif}, T.~H., {Springel}, V., \& {Bromm}, V. 2013, \mnras, 434, 3408

\bibitem[{{Greif} {et~al.}(2011){Greif}, {Springel}, {White}, {Glover},
  {Clark}, {Smith}, {Klessen}, \& {Bromm}}]{Greif+11a}
{Greif}, T.~H., {Springel}, V., {White}, S.~D.~M., {et~al.} 2011, \apj, 737, 75

\bibitem[{{Hartwig} {et~al.}(2015){Hartwig}, {Clark}, {Glover}, {Klessen}, \&
  {Sasaki}}]{Hartwig+15}
{Hartwig}, T., {Clark}, P.~C., {Glover}, S. C.~O., {Klessen}, R.~S., \&
  {Sasaki}, M. 2015, \apj, 799, 114

\bibitem[{{Hirano} {et~al.}(2014){Hirano}, {Hosokawa}, {Yoshida}, {Umeda},
  {Omukai}, {Chiaki}, \& {Yorke}}]{Hirano+14}
{Hirano}, S., {Hosokawa}, T., {Yoshida}, N., {et~al.} 2014, \apj, 781, 60

\bibitem[{{Hollenbach} \& {McKee}(1979)}]{Hollenbach+79}
{Hollenbach}, D., \& {McKee}, C.~F. 1979, \apjs, 41, 555

\bibitem[{{Hosokawa} {et~al.}(2016){Hosokawa}, {Hirano}, {Kuiper}, {Yorke},
  {Omukai}, \& {Yoshida}}]{Hosokawa+16}
{Hosokawa}, T., {Hirano}, S., {Kuiper}, R., {et~al.} 2016, \apj, 824, 119

\bibitem[{{Hosokawa} {et~al.}(2011){Hosokawa}, {Omukai}, {Yoshida}, \&
  {Yorke}}]{Hosokawa+11}
{Hosokawa}, T., {Omukai}, K., {Yoshida}, N., \& {Yorke}, H.~W. 2011, Science,
  334, 1250

\bibitem[{{Ishigaki} {et~al.}(2018){Ishigaki}, {Tominaga}, {Kobayashi}, \&
  {Nomoto}}]{Ishigaki+18}
{Ishigaki}, M.~N., {Tominaga}, N., {Kobayashi}, C., \& {Nomoto}, K. 2018, \apj,
  857, 46

\bibitem[{{Lau} \& {Bertin}(1978)}]{Lau+78}
{Lau}, Y.~Y., \& {Bertin}, G. 1978, \apj, 226, 508

\bibitem[{{Mayer} \& {Duschl}(2005)}]{Mayer+05}
{Mayer}, M., \& {Duschl}, W.~J. 2005, \mnras, 358, 614

\bibitem[{{McKee} \& {Tan}(2008)}]{McKee+08}
{McKee}, C.~F., \& {Tan}, J.~C. 2008, \apj, 681, 771

\bibitem[{{Omukai} \& {Yoshii}(2003)}]{Omukai+03b}
{Omukai}, K., \& {Yoshii}, Y. 2003, \apj, 599, 746

\bibitem[{{Ripamonti} \& {Abel}(2004)}]{Ripamonti+04}
{Ripamonti}, E., \& {Abel}, T. 2004, \mnras, 348, 1019

\bibitem[{{Schauer} {et~al.}(2019){Schauer}, {Liu}, \& {Bromm}}]{Schauer+19}
{Schauer}, A. T.~P., {Liu}, B., \& {Bromm}, V. 2019, \apjl, 877, L5

\bibitem[{{Shakura} \& {Sunyaev}(1973)}]{Shakura+73}
{Shakura}, N.~I., \& {Sunyaev}, R.~A. 1973, \aap, 24, 337

\bibitem[{{Sobolev}(1960)}]{Sobolev}
{Sobolev}, V.~V. 1960, {Moving envelopes of stars}

\bibitem[{{Stacy} {et~al.}(2016){Stacy}, {Bromm}, \& {Lee}}]{Stacy+16}
{Stacy}, A., {Bromm}, V., \& {Lee}, A.~T. 2016, \mnras, 462, 1307

\bibitem[{{Stacy} {et~al.}(2010){Stacy}, {Greif}, \& {Bromm}}]{Stacy+10}
{Stacy}, A., {Greif}, T.~H., \& {Bromm}, V. 2010, \mnras, 403, 45

\bibitem[{{Susa}(2019)}]{Susa+19}
{Susa}, H. 2019, \apj, 877, 99

\bibitem[{{Tan} \& {McKee}(2004)}]{Tan+04}
{Tan}, J.~C., \& {McKee}, C.~F. 2004, \apj, 603, 383

\bibitem[{{Tanaka} {et~al.}(2002){Tanaka}, {Takeuchi}, \& {Ward}}]{Tanaka+02}
{Tanaka}, H., {Takeuchi}, T., \& {Ward}, W.~R. 2002, \apj, 565, 1257

\bibitem[{{Turk} {et~al.}(2009){Turk}, {Abel}, \& {O'Shea}}]{Turk+09}
{Turk}, M.~J., {Abel}, T., \& {O'Shea}, B. 2009, Science, 325, 601

\bibitem[{{Yoshida} {et~al.}(2006){Yoshida}, {Omukai}, {Hernquist}, \&
  {Abel}}]{Yoshida+06}
{Yoshida}, N., {Omukai}, K., {Hernquist}, L., \& {Abel}, T. 2006, \apj, 652, 6

\end{thebibliography}

\end{document}